# Information-Centric Connectivity

Konstantinos V. Katsaros, Vasilis Sourlas, Ioannis Psaras , Sergi Reñé and George Pavlou
University College London, UK

**Abstract**

Mobile devices are often presented with multiple connectivity options usually making a selection either randomly or based on load/wireless conditions metrics, as is the case of current offloading schemes. In this paper we claim that link-layer connectivity can be associated with information-availability and in this respect connectivity decisions should be information-aware. This constitutes a next step for the Information-Centric Networking paradigm, realizing the concept of *Information-Centric Connectivity* (ICCON). We elaborate on different types of information availability and connectivity decisions in the context of ICCON, present specific use cases and discuss emerging opportunities, challenges and technical approaches. We illustrate the potential benefits of ICCON through preliminary simulation and numerical results in an example use case.

## 1 Introduction

Motivated by the proliferation of content-centric applications in the Internet, *Information-Centric Networking* (ICN) promises a shift in the operation of the network, enabling routing and forwarding based on identifiers/names of content, rather than network locations [1]. As such, ICN research efforts so far have primarily focused on changing/evolving the protocol stack at the network layer, and above, further introducing and investigating appealing features, such as the support for in-network caching, multicast and mobility.

At the same time, user demand for information is increasingly expressed through handheld devices (*User Equipment* (UE)) and ubiquitous wireless networks, as demonstrated by the substantial growth of mobile traffic [2]. Mobile devices are nowadays often presented with multiple connectivity options. Though typically associated with a single cellular operator, users are frequently in the vicinity of multiple IEEE 802.11 (WiFi) networks [3], usually offloading Internet traffic [4], but also increasingly providing access to locally stored information (*e.g.*, major airlines offer WiFi-based multimedia services [1]), or even enabling device-to-device (D2D) opportunistic communication between users [5], through WiFi Direct (*e.g.*, FireChat[2]).

As a result, users are presented with a multitude of opportunities to access information available in their networking vicinity in the form of opportunistically cached, pre-fetched/downloaded or locally generated (*e.g.*, UE) content and/or services/applications providing static or dynamically generated content. Depending on the networking environment, information can reside at a wide range of accessible network locations *i.e.*, UEs, WiFi APs with storage/processing capabilities, in-network content-centric routers, in-network caches / middle-boxes [6] or micro-clouds [7]. Spatiotemporal (*e.g.*, mobility) and application level user dynamics, as well as any explicit differentiation of accessible services and/or content, contribute to the expectation of information diversity in these locations. In turn, awareness of the availability (or expectancy) of information in the networking environment can lead to connectivity decisions that reflect user interests, thus facilitating or even enabling the otherwise impossible access to the desired information.

However, connectivity decisions are currently information agnostic. Offloading mechanisms are primarily focused on load/performance metrics such as downlink/uplink throughput, signal-to-noise ratio, *etc.* of WiFi connections [4], disregarding information availability. Moreover, searching for information currently builds on the assumption that a UE has already associated with a particular network device, either a WiFi AP or another UE. As a result, UEs need to engage in an iterative, time and energy consuming process comprised of the network association [8] and the subsequent search for information at the application level.

---

[1] *E.g.*, http://www.onair.aero/en/commercial-airlines-products
[2] https://en.wikipedia.org/wiki/FireChat

In this paper we argue that the ICN paradigm can and should be extended beyond the scope of the network layer (and above), enabling *information-centric connectivity* (ICCON). In ICCON, information-centrism is further expressed in the connectivity decisions taken by UEs, which aim at discovering the networks enabling or facilitating access to the desired information. In essence, information-awareness is introduced at the link layer[3]. In a characteristic ICCON use case, the selection of a WiFi AP for offloading cellular traffic can be driven by the matching between the user content interests and the currently available content at the AP and/or the broadband remote access router (BRAS), either opportunistically cached or pro-actively pushed there. In another example case, a WiFi SSID is carefully setup and selected by UEs to enable the exchange of information for a local event *e.g.*, photos taken during a concert.

ICCON is expected to enhance user experience as information-centric connectivity decisions bring the user closer to the desired information, reducing latencies, along with network traffic, *e.g.*, mobile network offloading, and server load. Information availability is further expected to improve when a connectivity decision either leads to the desired information or not *e.g.*, accessing photos in the aforementioned D2D example. Note that in the context of ICCON, these benefits come without the currently imposed need to search for information upon the time and energy consuming network association process. This comes in sharp contrast to a substantial body of work on service discovery, which, in most cases, assumes the establishment of connectivity between participating devices, before any service discovery protocol is employed (*e.g.*, Jini, UPnP). At the same time, the ability of UEs to intelligently discover information in their networking vicinity enables new opportunities for network operators, content and service providers, or even users themselves, to provide access to information with dedicated, low cost equipment (*e.g.*, APs, UEs), decoupling information provisioning from Internet access.

In the following, we elaborate on the ICCON concept thoroughly discussing the various connectivity options and decisions UEs can encounter, subject to different types of information available in their vicinity (Section 2.1 and 2.2). We elaborate on the mechanisms required to realize ICCON (Section 2.3) and further present two example use cases (Section 3). We identify a series of challenges to be met for the realization of the ICCON concept (Section 4) and we delve into the details of a particular offloading use case presenting preliminary results that demonstrate the potential benefits of ICCON (Section 5).

## 2 Information-awareness, connectivity options and decisions

In environments with rich connectivity options, ICCON employs information-awareness to enable the selection of the network/device that better facilitates or enables access to the information desired by the user. In the following, we elaborate on this concept by first identifying the connectivity options/scenarios of interest, along with the resulting types of information availability. Then, we elaborate on the exact objective of a connectivity decision in each identified scenario, paying particular attention on the resulting implications on the naming of information. Finally, we identify and discuss major aspects of the mechanisms envisioned to enable connectivity decisions in ICCON.

### 2.1 Connectivity options and information availability

In this work, we focus on the proliferating WiFi technology and the connectivity options enabled by IEEE 802.11 network interfaces, though the broader ICCON concept is not bound to a specific wireless technology. Bluetooth is also considered as an ICCON enabler, however we focus on WiFi technologies[4] as they support substantially longer ranges, they are currently dominant in providing wireless access to the Internet, and they support much faster device discovery [14].

---

[3] ICCON aims at supporting connectivity decisions per wireless network interface and as such, it does not focus on the concurrent use of multiple network interfaces, though it is fully compatible with such approaches e.g., [9]. Moreover, focusing on connectivity decisions, ICCON is orthogonal to higher layer mechanisms.

### 2.1.1 WiFi Hotspots

We first consider the case of WiFi hotspots enabling Internet access, through the access network of *Internet Service Providers* (ISPs). The proliferation of WiFi technology in this case is manifested by the often highly dense deployments of hotspots in urban environments *e.g.*, even up to several tens of APs [3]. In such environments, we first focus on information available in the form of content at the AP (*e.g.*, set-top-box with storage), at in-network middleboxes (*e.g.*, WAN accelerators) or content-centric routers. Content, in these cases, can be either opportunistically cached or pre-fetched in a *content distribution network* (CDN) fashion. Moreover, in view of the expected emergence of in-network processing capabilities [7], we further consider the availability of information in the form of low-latency, in-network services/applications *e.g.*, a local touristic guide offering route directions. We note that we consider information availability only loosely coupled with the provision of Internet access.

### 2.1.2 D2D Communication

Beyond WiFi hotspots, we are further inspired by scenarios where multiple users are co-located in crowded areas (*e.g.*, metro during rush hours, concerts, football games, *etc.*) presenting a rich set of connectivity opportunities. In such environments, access to the Internet is often not available (*e.g.*, underground metro), or rather limited (*e.g.*, flash-crowds deplete mobile network resources). Since most of the UEs are nowadays equipped with WiFi interfaces, the formation of WiFi Direct Groups (*i.e.*, D2D communication) appears as an appealing alternative option in the quest for information; especially when this information is linked to the concentration of the users (*e.g.*, photos during a concert). Such information can have the form of locally (user) generated or previously downloaded/retrieved content.

We note that connectivity decisions are not strictly bound to existing information, as the expectation of some information generated (or made available) in the future can also motivate the selection of a certain network. For instance, a WiFi Direct Group can be used in the aforementioned example, to setup a wireless broadcast domain for the delivery of photos taken during a concert. In such scenarios, a connectivity decision corresponds to the subscription to particular information, in a fashion similar to radio stations or TV channels.

## 2.2 Connectivity decisions and information naming

As the high level objective of ICCON is to facilitate access to the desired information, a connectivity decision goes through the comparison of user interests against the available/expected information. In the case of cached content the objective of ICCON is to identify the connectivity option that maximizes the matching with the users' interests, and in turn the likelihood of hitting the cache(s). To this end, UEs keep track of the requested content, essentially building a *UE profile* through time. Content is identified following the information naming conventions of the caches *e.g.*, URL, CCN/NDN chunk name. On the other side, a cache index provides a description of the content available forming a corresponding *network profile*. Similar profiling mechanisms can be envisioned for pre-fetched content or for services.

This matching mechanism can also support cases where a user is searching for a particular information item (*e.g.*, an image file), that is, when the objective of a connectivity decision is to facilitate both discovery and access to information. However, taking into account the enormous size of the information space, enabling the search for individual information items would likely result in significant search overheads, in some cases only to facilitate access to a single item, potentially not justifying a connectivity decision. We believe that avoiding these overheads goes through the appropriate structuring of the information namespace. The envisioned structure is designed around the notion of services/applications enabling the aggregation of individual information item names within their semantic scope *e.g.*, aggregating articles in a news service. Similarly to publish/subscribe systems, services/applications may further define individual topics for the support of finer grained semantic scopes *e.g.*, a topic for the World Cup Final can be created by a sports news service. ICCON then targets at discovering the desired services/applications (or topics therein) within each connectivity option. Finer-grained information discovery and retrieval is subsequently supported by UE-side applications *i.e.*, upon a connectivity decision. This avoids the need for a universal naming scheme of information items, promises better scalability, as only a few million applications are currently available at the most popular application market places, and builds on the currently prevailing application-centric usage model of handheld devices[5]. The exact representation of available services/applications (or topics therein) in this category heavily depends on the control plane mechanisms enabling ICCON. We elaborate on these mechanisms in the following.

---

[4] We refer to technologies following the IEEE 802.11 standard, as well as its IEEE 802.11u amendment[11], WiFi Direct and the recently announced WiFi-NAN protocol[12].

### 2.3 ICCON mechanisms

ICCON requires a control plane mechanism for the exchange and comparison of the representations of the desired and the available/expected information. Depending on the networking environment, we consider a series of aspects and corresponding design options for this mechanism.

**Push *vs.* pull.** Candidate WiFi APs or UEs can pro-actively (*push*) or re-actively (*pull*) advertise their available/expected information. Client UEs (*i.e.*, information consumers) may also have their profiles or service/application/topic identifiers pushed or pulled. The exact selection depends on aspects such as privacy concerns (*i.e.*, revealing user interests, see also Section 4), energy availability and incentives schemes (*e.g.*, energy footprint for advertising available content residing at an AP *vs.* a UE), the ratio of information providers over consumers, *etc*.

**Decision maker.** The comparison between the desired and the available/expected information can be made by different entities depending on the available connectivity options and the push/pull operation mode. In the case of mobile network offloading, the Access Network Discovery and Selection Function (ANDSF) [10] at the mobile network operator side can take this role, collecting the necessary information and reducing the associated overheads for UEs. However, privacy concerns motivate the placement of this decision process at the UE side, though we can envision profile representation schemes overcoming this limitation (see Section 4).

**Protocol layer and technical enablers.** Depending on the use case, the exchange of the aforementioned representations can take place at the link layer or above. In the case of mobile network offloading, network profiles can be refreshed frequently (periodically) at the ANDSF side, while UE profiles can be collected on demand (*i.e.*, upon an offloading decision) [10]. When UEs autonomously select their network of preference, a link layer mechanism is required to support the exchange of the representations. UEs can retrieve network profiles through the Access Network Query Protocol (ANQP) of IEEE 802.11u[6] *e.g.,* the *ANQP vendor-specific list* element [11]. The recently announced WiFi Neighbour Awareness Networking (NAN) protocol [12] also further supports a low energy consumption device discovery mechanism enhanced with publish/subscribe primitives that can serve the same purpose. The same link layer mechanisms can be used to support the exact matching of service/application/topic identifiers (see Section 2.2).

In the particular case of service/application/topic identifiers, we further consider the use of carefully selected non-human readable SSID values. Such values become available to UEs on the application layer, allowing only eligible UEs (and/or APs) to advertise or semantically link an SSID to a certain service/application/topic (see also Sections 2.2 and 4).

---

[5] We see the role of UE-side application components as particular important in further supporting access control mechanisms, reputation and incentives schemes for D2D communication. We do not further elaborate on these aspects due to length limitations.

[6] AQNP protocol builds on top of the Generic Advertisement Service (GAS) of IEEE 802.11u [11], which specifies a frame format and exchange process.

**Temporal granularity.** The control plane communication overhead heavily depends on the frequency of connectivity decisions. At the one extreme, we consider decisions to be made per content/service request. At the other extreme, decisions can be made once per UE, for a particular spatiotemporal context *e.g.*, set of available APs at a certain location and time. The selected granularity depends on the rate of user requests, the expected duration of the response to a request (*e.g.*, long YouTube video *vs.* a tweet) and the associated overheads of the control plane communication mechanism (see also Section 5).

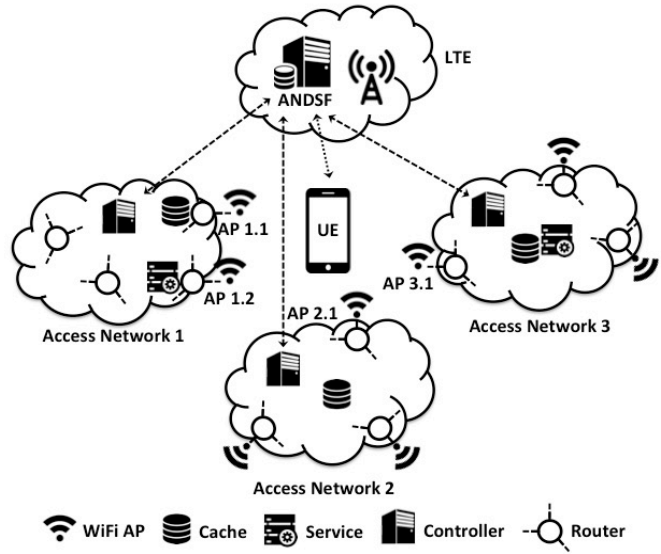

**Figure 1:** Mobile network operator supported offloading. Cached content or services may be collocated with the APs or reside further inside the access network (e.g., a centralised cache in Access Network 2). A local, per access network, controller reports information availability at the ANDSF component of the mobile network. A UE connects to one of the APs in its vicinity (*i.e.*, AP1.1, AP 1.2, AP 2.1 or AP3.1) based on recommendations from the ANDSF.

## 3 Use cases

As revealed in the previous section, a series of design choices result in a corresponding multitude of potential ICCON mechanisms. In this section we put together the available pieces providing a description of two example use cases to further illustrate ICCON functionality.

### 3.1 Mobile network operator supported offloading

We first focus on the abundance of WiFi networks [3] and the ICCON-enabled WiFi AP selection for offloading purposes. In our example, UE connectivity decisions are guided by the availability of the desired content in the caches of each available access network. UEs locally build their profiles by inspecting the requested URLs in the generated HTTP Requests (or the requested content names in Interest packets, in case of CCN/NDN networks [1]). A local *virtual* cache is used to track item popularity in a Least Frequent Used (LFU) fashion, for instance, without locally caching the content itself. On the access network side, the cache indexes are used as the network profiles.

As shown in Figure 1, the ANDSF component of the mobile network operator is responsible for collecting and comparing the UE and network profiles. To this end, a logically centralized controller established at each access network, is responsible for regularly collecting cache indexes and *pushing* them to the ANDSF. On the user side, Bob is in a café in a central square of the city, turns on the WiFi interface of his device and opens a video/music application (*e.g.,* YouTube, Spotify, *etc.*) to listen again to the latest album of his favorite singer. The UE pushes its profile to the ANDSF, which identifies an AP/SSID leading to a cache that holds a copy of the album. The UE receives a corresponding suggestion and subsequently proceeds with the association process (*e.g.*, with AP 1.1 in Figure 1). When the album finishes, Bobs proceeds to play another album triggering the same process, this time leading to the association with another AP (*e.g.*, with AP 3.1 in Figure 1).

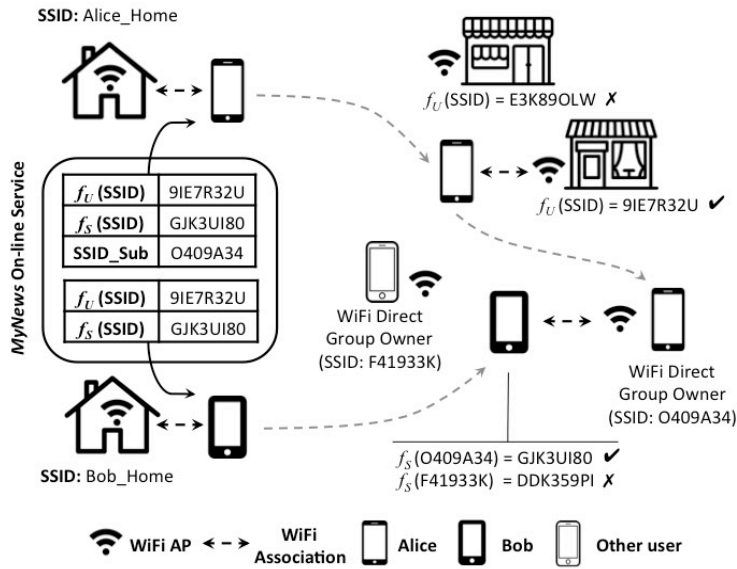

**Figure 2:** Autonomous connectivity decisions. The *MyNews* service provides Alice and Bob with functions $f_U$ and $f_S$ which enable them to identify SSIDs denoting the availability of news updates and other users' interest in music festival information respectively. Alice is also provided with an SSID value to be used for music festival information. Bob uses $f_S$ to identify Alice's subscription to *MyNews* festival information, denoted by the corresponding WiFi Direct Group. His UE joins the group to provide the festival program.

## 3.2 Autonomous connectivity decisions

In our second use case, we consider users looking for the latest news information, which is provided in the form of a service (see Section 2.2) *i.e.*, a news service *MyNews*. The content for this service is provided by a news agency, which pushes the news feeds to the on-line users of the corresponding smartphone application and certain ICCON-enabled WiFi APs around the city. The purpose of the UE-side application is to pull content when Internet access is available, and to guide the connectivity decisions of UEs in their quest for *MyNews* service feeds in ICCON-enabled environments.

Alice checks the news while at home. Later on, she roams in the city. As she has reached the quota of her cellular data plan she turns on the WiFi interface of her device and opens the *MyNews* application to get any updates. Her application has downloaded a set of SSID values (or their algorithmic properties) corresponding to WiFi-enabled *MyNews* repositories around the city. The ICCON-enabled WiFi manager of her device identifies one of the available SSIDs in one of the many beacons received from nearby APs and automatically connects using Alice's *MyNews* application user credentials. Alice selects the *Politics* category generating a request towards the AP, which responds with a data message (or more)[7].

At the same time, a set of SSIDs has been created to enable the sharing of user generated content, between *MyNews* users participating a big music festival. Alice goes to the festival and is looking for any related information. Her UE enters the *autonomous* group formation WiFi Direct mode [8], beaconing an SSID that can be recognized by other *MyNews* UEs. Bob is in the area and has previously downloaded the festival program using his *MyNews* application. His UE senses Alice's SSID and asks Bob if he is interested in sharing any information. Bob allows his UE to connect to

---

[7] Request/response messages may be realized on the application layer or by Interest/Data messages.

Alice's group and provide the festival program. The application credits Bob's account, enabling him in turn to search for information during a football match the following week.

## 4   Challenges

The description of the ICCON concept and the presented use cases, implicitly reveal a wide range of challenges and open issues to be addressed. In the following we identify and discuss some of these issues.

**Profiling.** In the mobile offloading case, the selection of an access network is made based on the match between the UE and the network information profiles. Since network profiles correspond to caches in the network, it is essential to consider the scope of the profiling as expressed by the number of caches taken into account, their location and size, in order to assess the potential benefits of guiding UE decisions. For instance, as shown in [6], large centralized caches are expected to yield higher cache hit ratios, but at the same time reduced cached item lifetimes, potentially rendering good profile matches practically useless (see also Section 5). Also, application level dynamics (*e.g.,* request rates) need to be considered in the selection of the temporal granularity of the matching mechanism, where connectivity decisions can, in certain cases, be taken on a per request basis.

At the same time, the representation of UE and network profiles presents a series of technical challenges. UE energy limitations, as well as performance reasons, impose important size limitations in the representation of UE profiles. Namely, profile comparison needs to be fast as to reduce (i) the corresponding energy consumption, when performed by a UE, (ii) the time required to take a connectivity decision. Set reconciliation methods (e.g., [13]) can be revisited in the ICCON context, as a potential approach towards lightweight profile matching. The use of Bloom filters can also assist in the compact and privacy conserving representation of profiles[8].

The creation of UE profiles also poses challenges in precisely reflecting user preferences. Simple content item popularity may be augmented with contextual information such as temporal characteristics (*e.g.*, time of the day), and social information (*e.g.*, different profiles when at work or with friends).

**Holistic connectivity management.** Information availability should be considered in the broader context of connectivity management, along with multiple other performance-related aspects, such as the wireless conditions (*e.g.*, signal-to-noise ratio), the available data rates, the observed load on the devices providing the available information (*e.g.*, caches). A careful fine-tuning of the weight of each aspect is obviously required. To this end, using naming to expose information about the traffic type of the desired content/service (*e.g.*, low latency interactive traffic *vs.* bulk download) can also assist in decision-making.

**Content/service placement.** In pull-based mechanisms (see Section 2.3), UEs may provide indications of the desired information implicitly allowing listening devices (*e.g.*, APs) to collect information about the desired information in certain areas. This information can be subsequently used to take decisions on the placement of the desired content/services in a networking area, taking further into account other parameters such as the availability of resources.

**Security and incentives.** Exposing UE profile information can raise privacy concerns. Carefully encoding user preferences is required, possibly calling for the cooperation between information providers and users *i.e.*, agreeing on the naming scheme and the encoding method. Enabling access to content/services through corresponding applications provides such a means. Moreover, the truth-full advertising of information must be ensured so as to guarantee the overall system utility. Cryptographic signatures, pre-shared private/public key pairs, as well as reputation-based mechanisms may contribute towards this direction. In the case of service-specific SSIDs (see Sections 2.2 and 3.2), efficient algorithms for the selection of the appropriate SSID values are required. Furthermore, appropriate countermeasures are required for DoS attacks targeting the overwhelming of UEs with advertised information. Finally, in D2D scenarios, users must be incentivised to offer their resources. Enabling communication through smart applications, also presents the potential for an out-of-band, credit management schemes linked to user application accounts.

---
[8] For instance, a 24KB Bloom filter can describe a cache index of $10^4$ items, with a false positive ratio of $10^{-4}$.

**Naming granularity and spectrum sharing.** Setting up per service/application/topic SSIDs relates information naming granularity to the use of the wireless spectrum. Subject also to information demand, *too* fine-grained naming can lead to overheads related to the support of high numbers of SSIDs, and broader medium access control. This calls for a closer look on the impact of naming granularity to the utilization of the wireless spectrum, and the sharing of wireless medium by the various services/applications/topics.

## 5 Preliminary results

Taking a first step in addressing some of the identifying challenges, we engage in a preliminary investigation of the potential benefits of ICCON in our first example use case *i.e.*, mobile operator supported offloading (see Section 3.1). To this end, using a custom Matlab-based simulator, we first explore the potential benefits in terms of cache hit ratio (CHR). Then, we investigate the impact of the different cache aggregation levels on the characteristic time of cache objects and its relation to the inter-arrival time of user requests (see Section 4).

### 5.1 Impact of ICCON on cache hit ratio

In our first setup, a set of *N* UEs affiliated with a single mobile operator visit a certain location where a set of *M* WiFi APs are visible, offering access to the Internet. Each AP is backed-up by a single cache of size *c*. Users request content from a catalogue of size *C*. Content popularity follows a Zipf-like distribution of slope *s*. *U* unique UE profiles are derived from the Zipf-like distribution. Each profile consists of a set of *u* unique, uniformly randomly selected items from the content catalogue.

In this context we first consider *N/3* of the users to be initially associated to a randomly selected AP, each generating content requests at a rate $\lambda_c$. Once caches have stabilized, an arrival-departure process starts. At each step, one already connected user departs from the system[9], and a new one enters it, connecting to the AP leading to the cache with the best fit against its profile. Between consecutive departure/arrival events we let caches stabilize. The process has a rate $\lambda_v$ and completes once all 2N/3 ICCON-supported UEs enter the system.

The selection of the AP for the ICCON-supported UEs is based on a fit function *F* that takes into account the weighted (*w*) average of both the UE-network profile match (*f*) and AP load (*l*) *i.e.*,

$$F = wf + (1-w)l$$

*f* is calculated as the ratio a UE profile's content items found in the LFU-index of the corresponding cache, and *l* is calculated as:

$$l = 1 - \frac{n_i}{\sum_{i=1}^{M} n_i}$$

where $n_i$ denotes the number of users currently being served at AP $i, i \in [1..M]$. The ANDSF component of the mobile operator calculates *F* on behalf of UEs (see Section 3.1). Figure 4a shows the evolution of the CHR observed across APs for the entire lifetime of the aforementioned departure/arrival process, comparing against the random selection of APs. Interestingly enough, we observe that when half of the initially randomly assigned UEs depart from the network (*i.e.*, 25 UEs here) the ICCON mechanism starts performing better than the random case, since from that point on, it manages to cluster the UEs at different APs based on their profile. ICCON increasingly improves the observed CHR, until it reaches an increase of 10% against the non-ICCON case.

---
[9] Here we consider a FIFO departure scheme.

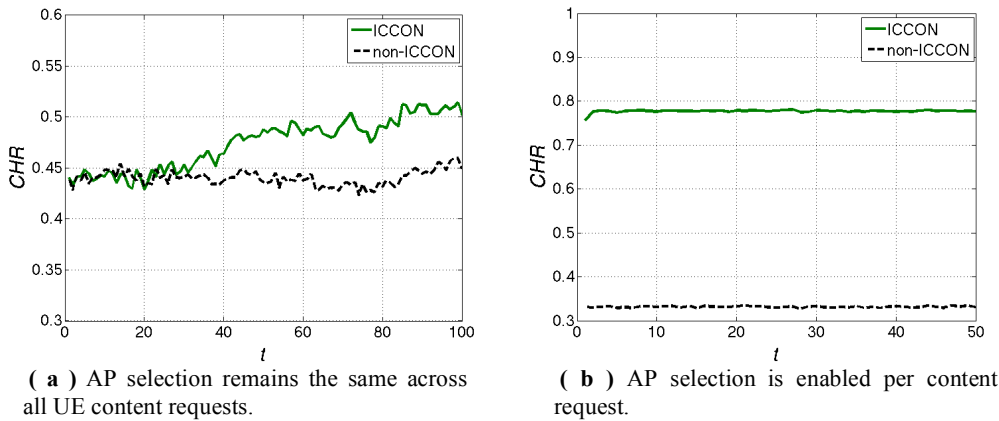

( a ) AP selection remains the same across all UE content requests.

( b ) AP selection is enabled per content request.

**Figure 3:** Impact of ICCON supported AP selection on CHR [$N$=150, $M$=10, $C$=$10^4$, $c$=5%$C$, $s$= 0.8, $\lambda_c$ =0.01 req/sec, $\lambda_v$=0.003 users/sec, $U$=50, $u$=10%$C$, $w$=0.65]. Time is measured in total number of arrival/departures.

In the same context, we further examine the impact of ICCON when connectivity decisions are taken per content request. Here we consider 50 time slots of 100K requests each, with each request delivered to the AP with the best $F$ value. Figure 3b shows the observed CHR. We see a significant improvement against the non-ICCON scenario, quickly reaching a CHR of around 78%.

## 5.2 Impact of cache aggregation level

We further investigate the impact of the size of the aggregated cache load on the practical usefulness of ICCON *i.e.*, we are interested in the impact of this load to the lifetime of items in a cache as compared to user request inter-arrival times. We express this aggregated load with the *aggregation level* parameter $\alpha$, which denotes the number of UEs sharing a single cache. Using Che's approximation [15] we compare the *characteristic time* ($\tau$) (*i.e.*, the time spent in the cache until evicted in a Least Recently Used scheme) of cached objects that match a certain UE profile against the inter-arrival time of UE requests ($1/\lambda_c$). For this purpose we define: $r=\tau\lambda_c$. Obviously, ICCON becomes meaningful for values greater than one *i.e.*, when cached items are more likely to not have been evicted before they get requested.

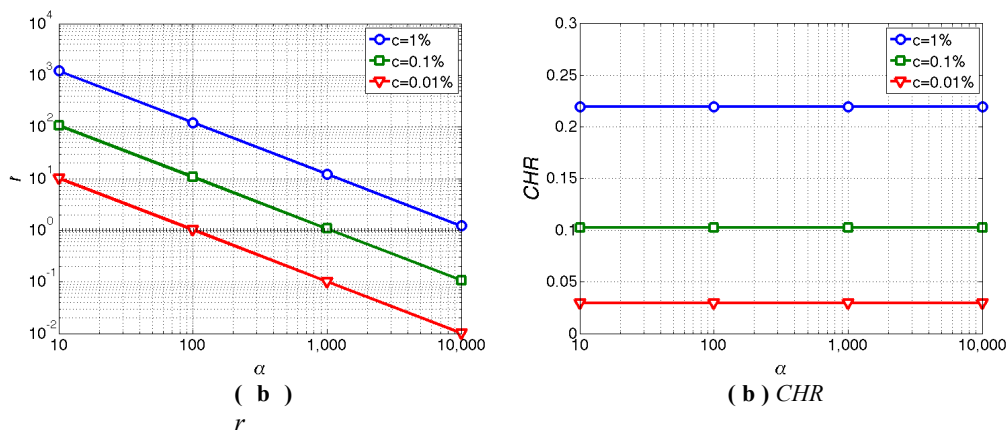

( b ) $r$

( b ) *CHR*

**Figure 4:** Impact of aggregation level $\alpha$ on $r$ (a) and the CHR (b). The results correspond to C=$10^6$ and $\lambda_c$=0.01requests/sec. Cache sizes ($c$) are expressed as a ratio of the catalogue size.

Figure 4 shows the $r$ and CHR values for different aggregation levels and cache sizes. We see that large caches and low UE populations significantly increase $r$ (note the logarithmic scale on both axes). We notice that an order of magnitude larger cache size has the same effect as an order of magnitude lower $\alpha$. Highly centralized caches result in low $r$ values, lowering the expectations from a good profile match. However, as expected, Figure 4b shows considerably lower CHR for small cache sizes, resulting in small UE benefits, even if content is actually found in a cache, as indicated by the profile matching.

# 6 Summary and Conclusions

In this article, we proposed the extension of the ICN paradigm to further encompass connectivity decisions, so as to enable the efficient discovery and access of the desired information in the networking vicinity of mobile devices. The resulting ICCON concept can be applied in several different environments ranging from cellular offloading decisions to D2D communications. Though several technical enablers have been identified, a series of research challenges need to be faced so as to rip the benefits of ICCON. Our preliminary investigations demonstrate such benefits in the example case of cellular offloading, motivating further research efforts.